\documentclass[preprint,5p,times,twocolumn]{elsarticle}
\usepackage{graphicx}
\usepackage{subfigure}
\usepackage{amssymb}
\usepackage{amsmath}
\usepackage[dvipsnames]{xcolor}
\journal{Physica D}
\begin{document}
\begin{frontmatter}
\title{Opinion dynamics involving contrarian and independence behaviors based on the Sznajd model with two-two and three-one agent interactions}
\author[brin]{Roni Muslim \corref{cp}}
\ead{muslim.roni@gmail.com}
\cortext[cp]{Corresponding author}
\author[brin]{M Jauhar Kholili}
\ead{m.jauhar.kholili@brin.go.id}
\author[brin]{Ahmad R. T. Nugraha}
\ead{ahmad.ridwan.tresna.nugraha@brin.go.id}
\affiliation[brin]{
    organization={Research Center for Quantum Physics,
    National Research and Innovation Agency (BRIN)},
    city={South Tangerang},
    postcode={15314},
    country={Indonesia}
}
\begin{abstract}
We investigate the opinion evolution of outflow dynamics based on the Sznajd model on a complete graph involving contrarian and independence behaviors. We consider a group of four spins representing the social agents with the following scenarios: (1) scenario two-two with contrarian agents or independence agents and (2) scenario three-one with contrarian or independence agents. All of them undergo a second-order phase transition according to our simulation. The critical point decreases exponentially as $\lambda$ and $f$ increases, where $\lambda$ and $f$ are contrarian and flexibility factors, respectively. Furthermore, we find that the critical point of scenario three-one is smaller than that of scenario two-two. For the same level of $\lambda$ and $f$, the critical point of the scenario involving independence is smaller than the scenario with contrarian agents. From a sociophysics point of view, we observe that scenario three-one can likely reach a stalemate situation rather than scenario two-two. Surprisingly, the scenarios involving contrarians have a higher probability of achieving a consensus than a scenario involving independence. Our estimates of the critical exponents indicate that the model is still in the same universality class as the mean-field Ising model.
\end{abstract}

\end{frontmatter}
\section{Introduction}
Over the past decade, science has been developing very rapidly, with connections between one branch and another forming in various fields.  Physicists specializing in nonlinear phenomena and statistical physics, for example, have attempted to apply the relevant concepts to understand social and political phenomena~\cite{castellano2009statistical,serge2016sociophysics, sen2014sociophysics, javarone2014network}.  This form of interdisciplinary science is well known as sociophysics~\cite{serge2016sociophysics}, in which there emerge some statistical physics features such as phase transition, scaling, and universality.  One of the most discussed topics in sociophysics is opinion dynamics~\cite{castellano2009statistical, serge2016sociophysics, sen2014sociophysics, stauffer2009encyclopedia}.  To understand and predict opinion dynamics, physicists have tried to correlate the microscopic-macroscopic phenomena in the physical system to the social system, e.g., collective phenomena in the macroscale with the individual behavior in the microscale~\cite{myer2013}.  Such correlation is similar to what physicists study in thermodynamics and statistical physics so that various models of opinion dynamics from physics' point of view have emerged, such as the celebrated Sznajd model~\cite{sznajd2000opinion}, the Galam model~\cite{galam2008sociophysics}, the voter model~\cite{liggett1985interacting}, the majority rule model~\cite{mobilia2003majority,galam2002minority,krapivsky2003dynamics}, and the Biswas-Sen model~\cite{biswas2009model}.  Most of the models have a ferromagnetic-like character which makes the system always homogeneous, i.e., all members in the system have the same opinion.  The ferromagnetic-like character in those models represents conformity behavior in social literature~\cite{nail2000proposal}; however, the models are inadequate when faced with social reality.  Therefore, physicists have further been proposing several social parameters, such as contrarian~\cite{galam2004contrarian}, inflexibility~\cite{galam2007role}, nonconformity~\cite{nyczka2013anticonformity}, and fanaticism~\cite{mobilia2003does}, to make the models more realistic.

Inspired by social psychology, one can classify several social responses or behaviors such as conformity and nonconformity\cite{nail2000proposal, willis1963two, willis1965conformity, macdonald2004expanding, nail2011development}. Conformity is a behavior that obeys group norms, while nonconformity does not follow the group norms. Nonconformity further splits into two types, namely anticonformity and independence. Anticonformity is a behavior contrary to that of the group majority. Galam defines anticonformity as ``contrarian" behavior, in which an agent or individual takes a particular choice opposite to the group choices with a certain probability~\cite{galam2004contrarian}. Nonconformity agents play an important role in the dynamics, in which the conformity agents can give the stabilization in a complete graph, yet the opposite occurs in the incomplete graph~\cite{javarone2014social}. TThe density of conformity agents also plays an important role in opinion formation within the system. With a certain density of conformity agents, the system reaches a full consensus with all agents having the same opinion~\cite{javarone2015conformism}.

Implementation of the contrarian behavior in the outflow dynamics within the Sznajd model (``united we stand, divided we fall") exhibits an order-disorder nonequilibrium phase transition. The contrarian term in opinion dynamics is analogous with thermal fluctuation in the thermodynamic system, which acts to make the system disorder. It has been shown that the parameter has an impact on the system that is indicated by the presence of a nonequilibrium phase transition. With a certain threshold probability of contrarian $p_c$, the system reaches a stalemate situation with no majority-minority opinion~\cite{de2005spontaneous}. The same effect of anticonformity behavior occurs within the Sznajd model; with a certain probability $p_c$,  the system undergoes a second-order phase transition. In the social context, below the critical point $p_c$ there is consensus with all agents having the same opinion or the emergence of majority opinion~\cite{calvelli2019phase, muslim2020phase}.

The other subset of nonconformity, i.e., independence, is a behavior that involves an individual who is not following group norms but acting independently to the group norms. In the physics context, the behavior of independence is similar to the temperature in the Ising system; it acts like sources of noises that can induce a system to undergo an order-disorder phase transition~\cite{nyczka2013anticonformity, calvelli2019phase, sznajd2011phase, crokidakis2015inflexibility}. In addition, the presence of independence behavior adopted by agents in various scenarios causes the system to undergo a discontinuous phase transition, especially in the $q$-voter model~\cite{sznajd2011phase, chmiel2015phase, abramiuk2019independence, nowak2021discontinuous, civitarese2021external}.

The level of conformity or nonconformity is not the same in every community group; it depends on several factors such as culture, education, and age. The level of independence also varies in each country. One way to quantify the level of independence is using the individualism index value or the individualism distance index (IDV) proposed by Hofstede~\cite{hofstede2001, hofstede2010}. The IDV indicates a degree of thinking of yourself first as an individual and then considering the group after. The US (IDV = 92) and UK (IDV = 89) have higher IDV levels than Asian countries or ``conservative" countries, for example Indonesia (IDV = 14) and South Korea (IDV = 18). Although individualism is not the same as independence, we may observe both individualism and independence more commonly in countries with high IDV~\cite{solomon2010consumer}. The IDV data for other countries are available on the internet~\cite{hofstedeIDV}. 

The implementation of the independence behavior in the Sznajd model can be found in Ref.~\cite{sznajd2011phase}. In that paper, Sznajd-Weron et al. introduced a flexibility factor $f$ and analyzed the model on the lattice and a complete graph with three-spin (representing three-agent) interactions. On the other hand, the implementation of the independence behavior to the other opinion dynamics model can be found in Refs.~\cite{nyczka2013anticonformity, crokidakis2015inflexibility, abramiuk2019independence, nyczka2012phase}. However, there were no studies that combine and compare the contrarian behavior with independence based on the Sznajd model, especially for the interactions of the four agents with two-two agent interaction and three-one agent interaction.

In this paper, we investigate the opinion dynamics within the Sznajd model involving both behaviors of contrarian and independence. We consider four-spin (or four-agent) local interactions defined on a complete graph, as described in Section 2. We then divide the model into the so-called two-two scenario and the three-one scenario for developing the microscopic rules of the outflow dynamics in the presence of contrarian and independent characters. The two-two (three-one) scenario occurs when two (three) agents influence the other two (one) wherever they are. For the independence behavior, we use the flexibility parameter $f$ that describes the level of independence, as previously defined by Sznajd-Weron et al.~\cite{sznajd2011phase}. For the contrarian behavior, we introduce a contrarian factor $\lambda$ defining the level of contrarian character, i.e., how often an individual takes a particular choice opposite to that of the group majority. Based on the proposed model, we present the analytical and numerical results in Section 3. In particular, we focus on discussing the phase transition and universality of the model. Finally, in Section~\ref {sec.4}, we give our conclusions and outlook.

\section{Model and methods}
\label{sec.2}

\begin{figure}[ht!]
	\centering
	\includegraphics[width=5 cm]{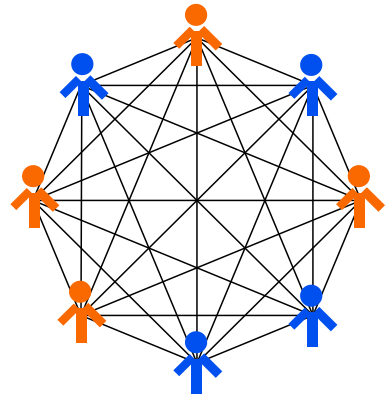}
	\caption{Illustration of a complete graph with 8 nodes (agents) with two different states (opinions) and 7 connection/edges (social relation in social context) for each node. This topology can be interpreted as a group interaction with each member of the group can interact each other. We can also consider all members of the group are neighboring each other.
	\label{cg}}
\end{figure}

We consider the outflow dynamics under the Sznajd model with contrarian and independence behaviors for four-agent local interaction. The model is defined on a complete graph (Fig. \ref{cg}) with total population $N = N_+ + N_-$, where $N_+$ and $N_-$ are total spin-up and spin-down states, respectively. The opinions of agents are represented by Ising spins  $S_n = \pm 1 \,(n = 1,2,\cdots, N)$. The initial condition of the system is $50\%$ up and $50\%$ down, i.e. the initial magnetization is zero (disorder state). We categorize the system of four-agent local interaction into (1) scenario two-two, where two agents persuade the other two in the population wherever they are, and (2) scenario three-one, where three agents persuade the fourth agent in the population, wherever it is in the population.

The macroscopic behavior of the system can be analyzed by considering the magnetization or the order parameter $m$, which, in the sociophysics point of view, is equivalent to the average opinion. In the analytical  consideration, the order parameter $m$ is defined as $m = 1/N \sum_{i=1}^{N} S_i$.  In the numerical simulation, the order parameter $m$ is an average of total samples, defined as $\langle m \rangle = 1/K \sum_{i =1}^{K} m_i$, where $\langle \cdots \rangle$ is the average of all samples, $K$ is the total number of samples, and $m_i$ is the order parameter of the $i$th sample. To estimate the critical exponents of the systems, we consider the ``magnetic" susceptibility $\chi$ and the Binder cumulant $U$, respectively defined as \cite{binder1981finite}:
\begin{align}
	\chi =& N \left( \langle m^2 \rangle - \langle m \rangle^2\right), \label{eq2}\\
	U =& 1 - \dfrac{\langle m^4 \rangle}{3\, \langle m^2 \rangle^2}. \label{eq3}
\end{align}
The universality class of the model can be defined by estimating the critical exponents of the model using finite-size scaling relations, defined as follows~\cite{cardy1996scaling}:
\begin{align}
	m ( N) & \sim N^{-\beta/\nu}, \label{eq4} \\
	\chi(N) & \sim N^{\gamma/\nu}, \label{eq5}\\
	U(N) &\sim \text{constant}, \label{eq6}\\
	p_{c}(N)-p_{c} & \sim N^{-1/\nu}, \label{eq7}
\end{align}
all of which are relevant around the critical point. We perform analytical calculations and numerical simulations and compare the results. The description of all scenarios of the model is given at each of the following subsections.

\subsection{Scenario two-two}
\label{subsec:2.1}
In this scenario, four agents represented by four spins are chosen randomly, namely, $S_i, S_j, S_k,$ and $S_l$, where two agents persuade the other two wherever they are in the populations according to the following microscopic rules:
\begin{itemize}
	\item For the scenario  with  contrarian 
	\begin{itemize}
		\item  With probability $\lambda p$, agents with states $S_k$ and $S_l$  take opposite state/opinion to the state/opinion of agents $S_i$ and $S_j$ only if the four agents have the same opinion. This is the contrarian rule of the model, where  symbolically two configurations of the agents' opinion satisfy the rule as illustrated below:
        \begin{equation}\label{contrarian-lambdap}
        	\begin{aligned}
        		& (1) \, \uparrow\uparrow \Uparrow\Uparrow \quad  \rightarrow \quad \uparrow \uparrow \Downarrow \Downarrow \\
        		& (2) \,\downarrow \downarrow \Downarrow \Downarrow \quad  \rightarrow \quad \downarrow \downarrow \Uparrow \Uparrow.
        	\end{aligned}
        \end{equation}
		\item With probability $(1-p)$, agents with states $S_k$ and $S_l$ follow the state/opinion of agents $S_i(t)$ and $S_j$ if $S_i = S_j$. Based on this rule, there are six configurations satisfying the rule as follows:
		\begin{equation}\label{contrarian_1-p}
			\begin{aligned}
				& (1) \,\uparrow \uparrow \Downarrow \Uparrow \quad  \rightarrow \quad \uparrow \uparrow \Uparrow \Uparrow \\
				& (2) \,\uparrow \uparrow \Uparrow \Downarrow \quad  \rightarrow \quad \uparrow \uparrow \Uparrow \Uparrow \\
				&\qquad \vdots\\
				& (6)\, \downarrow \downarrow \Uparrow \Downarrow \quad  \rightarrow \quad  \downarrow \downarrow \Downarrow \Downarrow.
			\end{aligned}
		\end{equation}
		This behavior is similar to the original Sznajd model (social validation)~\cite{sznajd2000opinion}, where for $p=0$, this model reduces to the original Sznajd model.
	\end{itemize}
	\item For the scenario with independence
	\begin{itemize}
		\item With probability $p$, agents $S_k$ and $S_l$ act independently. In other words, they do not follow the group norm; with probability $f$, the agents change, i.e., $S_{k,l}(t) = -S_{k,l}(t+dt)$, and with probability $(1-f)$, nothing changes, i.e., $S_{k,l}(t) = S_{k,l}(t+dt)$. The scheme of the agents configuration is illustrated as follows: 
		\begin{equation}\label{independent}
			\begin{aligned}
				& (1)\,\cdot \cdot \Uparrow \Uparrow \quad \rightarrow \quad \cdot \cdot \Downarrow \Downarrow \\
				& (2) \,\cdot \cdot \Downarrow \Downarrow \quad \rightarrow \quad \cdot \cdot \Uparrow \Uparrow.
			\end{aligned}
		\end{equation}
	The contrarian rule for the independence scenario is the same as ~\eqref{contrarian_1-p}.
	\end{itemize}
\end{itemize}

The symbols $\uparrow$ and $\downarrow$ stand for agents with states $S_i$ and $S_j$, respectively, while the symbols $\Uparrow$ and $\Downarrow$ stand for agents $S_k$ and $S_l$, respectively. The parameters $\lambda \in [0,1]$ and $f \in [0,1]$ are the contrarian factor and flexibility, respectively, which describe how often agents change their opinions in the contrarian and independence cases. These parameters are also similar to the stochastic driving parameter that causes the system to undergo an order-disorder phase transition \cite{nyczka2012phase, crokidakis2014phase}. 

Based on this model, there are sixteen agent combinations. Eight combinations are active, where six of which follow the conformity rule [illustrated in Eq.~\eqref{contrarian_1-p}] and two other combinations follow the contrarian [Eq.~\eqref{contrarian-lambdap}] or independence [Eq.~\eqref{independent}] rules. Eight other combinations are inactive because the polarity of agents persuader ($S_i$ and $S_j$) are tied (no or less contribution). From the social point of view, the inactive combinations can be interpreted as a weak opinion to persuade others. Therefore they do not affect the public opinion $m$ or the critical point of the systems.

\subsection{Scenario three-one}
\label{subsec:2.2}

In this scenario, four agents are chosen randomly, namely  $S_i,S_j, S_k$ and $S_l$, where three agents persuade the fourth agents, namely,  $S_l$  wherever it is in the population according to the following microscopic rules: 

\begin{itemize}
	\item For the scenario with contrarian
	\begin{itemize}
		\item With probability $\lambda p$, agent $S_l$ takes the opposite  state (opinion) to the majority state of the agents $S_i,S_j$ and $S_k$. Based on this rule, there are eight agent combinations satisfying the rule, as illustrated below:
		\begin{equation}\label{eq.confor31}
			\begin{aligned}
				&(1) \, \uparrow \uparrow \uparrow \Uparrow \quad \rightarrow \quad \uparrow \uparrow \uparrow \Downarrow \\
				&(2)\, \uparrow \uparrow \downarrow \Uparrow \quad \rightarrow \quad \uparrow \uparrow \downarrow \Downarrow \\
				& \qquad \vdots \\
				& (8) \, \downarrow \downarrow \uparrow \Downarrow  \quad \rightarrow \quad \downarrow \downarrow \uparrow \Uparrow
			\end{aligned}
		\end{equation}
		\item With probability $(1-p)$, agent $S_l$ follows the majority state of the agents $S_i,S_j$ and $S_k$. Based on this rule, there are eight agent combinations satisfying the rule. In this case, for $p = 0$, this model reduces to the original Sznajd model~\cite{sznajd2000opinion}.
		\begin{equation}\label{eq.contra31}
			\begin{aligned}
				& (1) \, \uparrow \uparrow \uparrow \Downarrow \quad \rightarrow \quad \uparrow \uparrow \uparrow \Downarrow \\
				& (2) \, \downarrow \uparrow \uparrow \Downarrow \quad \rightarrow \quad \downarrow \uparrow \uparrow \Uparrow \\
				& \qquad \vdots \\
				& (8) \, \uparrow \downarrow \uparrow \Downarrow \quad \rightarrow \quad \uparrow \downarrow \uparrow \Uparrow.			
			\end{aligned}
		\end{equation}
	\end{itemize}
	\item For the scenario with independence
	\begin{itemize}
		\item With probability $p$, the agent $S_l$ acts independently to the agents $S_i,S_j$ and $S_k$. In this case, with probability $f$, agent $S_l$ changes or flips, $S_l(t)= - S_l(t+dt)$, and with probability $(1-f)$ nothing changes,  $S_l(t)= S_l(t+dt)$ that illustrated as follows:
		\begin{equation}\label{eq.indep31}
			\begin{aligned}
				& (1) \cdots \Uparrow \quad \rightarrow \quad \cdots \Downarrow \\
				& \qquad \vdots \\
				& (8) \cdots \Downarrow \quad \rightarrow \quad \cdots \Uparrow.
			\end{aligned}
		\end{equation}
	Eight other spin combinations follow the conformity rule as illustrated in Eq.~\eqref{eq.confor31}.
	\end{itemize}
\end{itemize}

The symbols $\uparrow,\downarrow$ stand for agents $S_i,S_j,$ and $S_k$, while $\Uparrow,\Downarrow$ stand for agent $S_l$. If we consider the persuaders have the same state (opinion), the model reduces to the $q$-voter model with $q = 3$ and exhibits second-order phase transition~\cite{abramiuk2019independence}.

\section{Results and discussion}
\label{sec.3}
\subsection{Phase Transition}
In this section, we find the critical point that makes the system undergo an order-disorder phase transition. If we define the order parameter of the system as: 
\begin{equation} \label{order-param}
	m = \dfrac{1}{N} \sum_{i =1}^{N} S_i = \dfrac{1}{N} \left(N_{\uparrow} -N_{\downarrow}\right),
\end{equation}
and the concentration of spin up  is $c = N_{\uparrow}/N$, then from two relations above, we have $c = \left( m+1 \right)/2$, where $N_{\uparrow}$ and $N_{\downarrow}$ are total spin up and spin down respectively. During the dynamics process, total spins up $N_{\uparrow}$ increase by $1$, decrease by $-1$, or remain constant in a time step $t$. The value of $c(t)$ also simultaneously increases or decreases by $1/N$ or remains constant with the probabilities:
\begin{equation}\label{gamma-probs}
	\begin{aligned}
		\gamma_{+}  & = \text{prob}\left(c \rightarrow c+ 1/N\right), \\ 
		\gamma_{-} & = \text{prob}\left(c \rightarrow c- 1/N\right), \\
	\end{aligned}
\end{equation}
where the explicit formulations of $\gamma_{+}$ and $\gamma_{-}$ are given in each following system.

\subsubsection{Scenario two-two with contrarian agents}
\label{subsub.3.1.1}

\begin{figure}[t]
	\centering
  	\includegraphics[width=8.5cm]{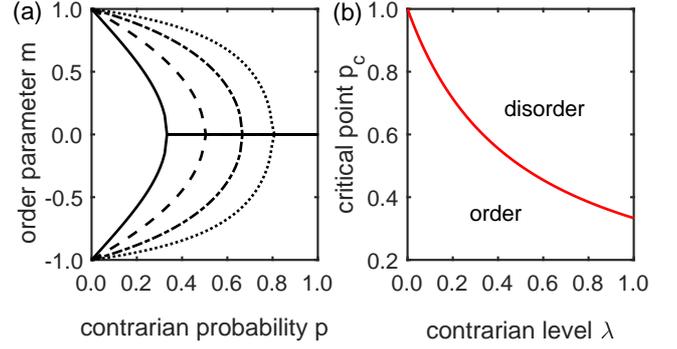}
	\caption{(a) Order parameter $m$ versus contrarian probability $p$ showing second-order phase transitions for the scenario two-two with contrarian agents [Eq.~\eqref{order-parameter2nd}] using typical values of contrarian level $\lambda$ (solid line: $\lambda = 1$, dashed line: $\lambda= 1/2$, dash-dotted line: $\lambda= 1/4$, dotted line: $\lambda= 1/8$). (b) Critical point $p_c$ versus contrarian level $\lambda$ according to Eq.~\eqref{critical-point22contra}.  As contrarian level $\lambda$ increases, the critical point $p_c$ decreases. This means that in a society with high level of contrarian (non conservative society), the probability to reach a stalemate situation is high. Equation~\eqref{critical-point22contra} also separates the order and disorder phases of the system.}
	\label{fig:22contra}
\end{figure}

For scenario two-two involving the contrarian agents, based on the model, there are eight active agent combinations: four combinations in which the spins flip from up-state to down-state and four other combinations, the spins (agents) flip from down-state to up-state. Therefore, the transition probability density of spin up increases $\gamma_{+}$ and decreases $\gamma_{-}$ are
\begin{equation}\label{gamma_prob2_contra2nd}
	\begin{aligned}
		{\it \gamma_{+}(c)} = & \,2\, \left( 1-p \right) {c}^{3} \left( 1-c \right) +2\, \left( 1-p \right) {c}^{2} \left( 1-c \right) ^{2} \\
		& +2\,\lambda\,p \left( 1-c \right) ^{4}, \\
		{\it \gamma_{-}(c)} = & \, 2\, \left( 1-p \right) c \left( 1-c \right) ^{3}+2\, \left( 1
		-p \right) {c}^{2} \left( 1-c \right) ^{2} \\
		&+2\,\lambda\,p\,{c}^{4},	
	\end{aligned}
\end{equation}
and by using the master equation that describes the "gain-lose" of the concentration $c$, we can write~\cite{krapivsky2010kinetic}:
\begin{equation}\label{gain-lose}
	\dfrac{dc}{dt} = \gamma_{+}(c) -\gamma_{-}(c).
\end{equation}
From Eqs.~\eqref{gamma_prob2_contra2nd} and~\eqref{gain-lose} for a stationary condition, $\gamma_{+}(c) =\gamma_{-}(c)$, we found for non zero $c$:
\begin{equation}\label{concentration_non-zero}
	 c = \dfrac{1}{2}\left(1\pm \left(\dfrac{1-\left(2\lambda +1 \right)p}{1+\left(2\lambda-1\right)p}\right)^{1/2}\right),
\end{equation}
and from the relation $m = 2\,c-1$, we found the order parameter $m$ that depends on the $p$ and $\lambda$:
\begin{equation}\label{order-parameter2nd}
	m = \pm \left(\dfrac{1-\left(2\lambda +1 \right)p}{1+\left(2\lambda-1\right)p}\right)^{1/2}.
\end{equation}
In this case, by setting $m = 0$, the critical point that makes the model undergo order-disorder phase transition depends on the level of contrarian $\lambda$, that is
\begin{equation}\label{critical-point22contra}
	p_c = \dfrac{1}{1+2\, \lambda},
\end{equation}
which shows that there are several second-order phase transitions for all values of contrarian level $\lambda >0$. We show plot of Eq.~\eqref{order-parameter2nd} for typical values of $\lambda$ in Fig.~\ref{fig:22contra}(a). From a sociophysics point of view, there is a majority opinion for $p < p_c(\lambda)$. In other words, when the contrarian probability $p = 0$ (there is no noise), there is a complete consensus with all members having the same opinion (ferromagnetic-like or completely ordered). The consensus level decreases for the probability of conformity $p(\lambda)$ increases until it reaches zero at the critical point $p_c(\lambda)$. Above the critical point $p_c(\lambda)$ there is no majority opinion or a stalemate situation (antiferromagnetic-like or complete disordered). Based on this model, one can also say, if the contrarian level is small (high conformity level), commonly found in conservative societies, the probability of the societies undergoing the status quo is small, and vice versa. It is also shown that, based on Eq.~\eqref{critical-point22contra}, the critical point $p_c$ decreases exponentially as $\lambda$ increases as exhibited in Figure \ref{fig:22contra}(b). Equation~\eqref{critical-point22contra} also separates into order and disorder phases.

\subsubsection{Scenario two-two with independence agents}
\begin{figure}[t]
	\centering
  	\includegraphics[width=8.5cm]{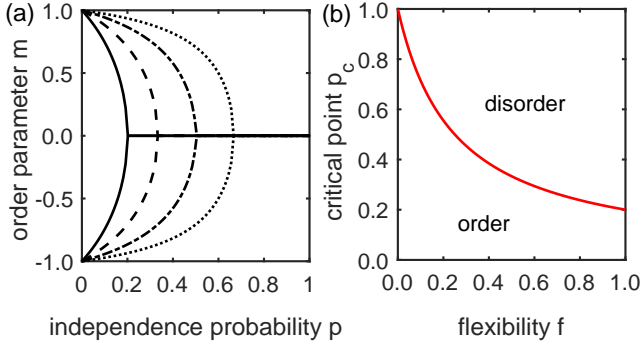}
	\caption{(a) Order parameter $m$ versus independence probability $p$ for the scenario two-two with independence agents [Eq. \eqref{order-param2_indep}] using typical values of flexibility $f$ (solid line: $f = 1$, dashed line: $f= 1/2$, dash-dotted line: $f= 1/4$, and dotted line: $f= 1/8$). (b) Critical point $p_c$ versus flexibility $f$ according to Eq.~\eqref{critical-point_indep22}. As flexibility $f$ increases, the critical point $p_c$ decreases. This result indicates that, in a society with high IDV, the probability to reach a stalemate situation is high.}
	\label{fig.22indep}
\end{figure}

For scenario two-two involving independence agents, six agent (spin) combinations follow the conformity rule (illustration \eqref{contrarian_1-p}), and two agent combinations follow the independence rule (illustration \eqref{independent}). Four agent combinations make agent change their opinion/state from 'up' to 'down', and four other agent combinations change their opinion/state from 'down' to 'up'. Therefore, the probability density of agent-up state increases $\gamma_{+}$ and decreases $\gamma_{-}$ can be written as follows:
\begin{equation}\label{gamma2_independent}
	\begin{aligned}
		\gamma_{+}(c) =& \, 2\, \left( 1-p \right) {c}^{3} \left( 1-c \right) +2\, \left( 1-p
		\right) {c}^{2} \left( 1-c \right) ^{2}\\
		&+2\,f \, p \,\left( 1-c \right), \\
		\gamma_{-}(c)  =& \, 2\, \left( 1-p \right) c \left( 1-c \right) ^{3}+2\, \left( 1-p
		\right) {c}^{2} \left( 1-c \right) ^{2}\\
		&+2\,f\,p\,c,
	\end{aligned}
\end{equation}
and for a stationary condition $\gamma_{+} = \gamma_{-}$. By using the relation $m = 2\,c-1$, the order parameter $m$ is given by
\begin{equation}\label{order-param2_indep}
	m = \pm \left(\dfrac{1-\left(1+4\,f\right)p}{1-p}\right)^{1/2},
\end{equation}
which shows the order parameter $m$ that depends on the flexibility $f$. Plot of Eq.~\eqref{order-param2_indep} for typical values of $f$ is depicted in Fig.~\ref{fig.22indep}(a). Based on this scenario, we also found that the critical point decreases exponentially as $f$ increases:
\begin{equation}\label{critical-point_indep22}
	p_c = \dfrac{1}{1+4\,f},
\end{equation}
which indicates that there are several order-disorder phase transitions for all values of $f > 0$. From sociophysics point of view, there is a coexistence of a majority-minority opinion for $p < p_c(f)$, and stalemate situation for $p>p_c(f)$, for all values of $f > 0$. Furthermore, if $f$ is small, the probability to reach a stalemate situation is high, and vice versa. The critical point $p_c$ in Eq.~\eqref{critical-point_indep22} separates the order-disorder phase as shown in Fig.~\ref{fig.22indep}(b).

Based on Eqs.~\eqref{critical-point22contra} and \eqref{critical-point_indep22}, the critical point for scenario two-two with independence agents is smaller than the scenario two-two with contrarian agents for the similar value of $\lambda$ and $f$ (the same level of contrarian and flexibility). It means that the probability to undergo an order-disorder phase transition for scenario two-two with agents of independent behavior is higher than scenario two-two with contrarian behavior. This is because the effect of the independence behavior makes the system is more 'chaotic' than the effect of the contrarian behavior in the system, thus, making the system more disordered. From a social point of view, the probability that reaches a status quo or a stalemate situation is more possible or more common in a society with independent behavior than in a society with contrarian behavior.

\subsubsection{Scenario three-one with contrarian agents}

\begin{figure}[t]
	\centering
    \includegraphics[width=8.5cm]{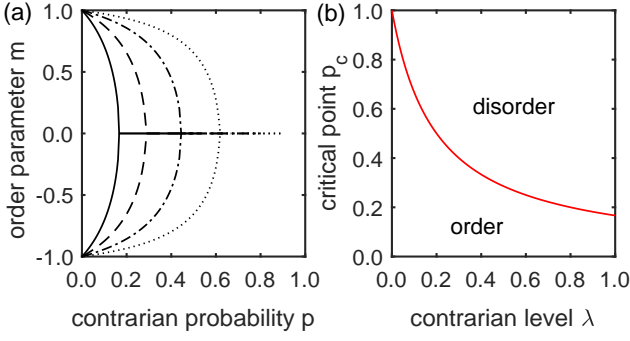}
	\caption{(a) Order parameter $m$ versus contrarian probability $p$ showing second-order phase transitions for the scenario three-one with contrarian agents [Eq.~\eqref{order-parameter3_contra}] for typical values of contrarian level $\lambda > 0$ (solid line: $\lambda = 1$, dashed line: $\lambda= 1/2$, dash-dotted line: $\lambda= 1/4$, dotted line: $\lambda= 1/8$). (b) Critical point $p_c$ versus contrarian level $\lambda$ according to Eq.~\eqref{eq.critical31contra}.  As contrarian level $\lambda$ increases, the critical point decreases exponentially and it also separates the order and disorder phases of the system. }
	\label{fig.31contra}
\end{figure}

For scenario three-one, three agents persuade the fourth agent. There are sixteen active agent combinations, where eight agent combinations follow the conformity rule [Eq.~\eqref{eq.confor31}] and eight other agent combinations follow the contrarian rule [Eq.~\eqref{eq.contra31}]. Eight agent combinations change their opinion/state from 'up' to 'down', and vice versa. Therefore, the probability density of agent-up state increases and decreases can be written explicitly as:
\begin{equation}\label{eq.gamma31_2nd}
	\begin{aligned}
		\gamma_{+}(c) = & \, {c}^{3} \left( 1-c \right)  \left( 1-p \right) +3\,{c}^{2} \left( 1-c \right) ^{2} \left( 1-p \right) \\ 
		& +3\,c \left( 1-c \right) ^{3}\lambda \,p+ \left( 1-c \right)^{4}\lambda\,p, \\
		\gamma_{-}(c)  = & \, c\,\left( 1-c \right) ^{3} \left( 1-p \right) +3\,{c}^{3} \left( 1-c \right) \lambda\,p \\ 
		& +3\,{c}^{2} \left( 1-c\right) ^{2} \left( 1-p \right)  +{c}^{4}\lambda\,p,
	\end{aligned}
\end{equation}
where $p_1 = 1-p $ and $p_2 = \lambda p$ are the conformity and contrarian probability, respectively. For a stationary state, by using the relation $m = 2\,c - 1$, the order parameter $m$ of the system is found to be:
\begin{equation}\label{order-parameter3_contra}
	m =  \pm \left(\dfrac{1-\left(1+5\,\lambda\right)p}{1-\left(1+\lambda\right)p}\right)^{1/2},
\end{equation}
which depends on the level of contrarian $\lambda$ as shown in Figure ~\ref{fig.31contra}(a). The situation is similar to scenario two-two with contrarian agents, i.e., for small $\lambda$ the probability to reach the stalemate situation is high, and vice versa. If we compare scenario three-one versus two-two with contrarian agents, the critical point $p_c$ of scenario three-one is smaller than scenario two-two for the same level of $\lambda$.

The critical point $p_c$ of this scenario is given by
\begin{equation}\label{eq.critical31contra}
	p_c = \dfrac{1}{1+5\, \lambda},
\end{equation}
where the critical point $p_c$ decreases exponentially as the contrarian level $\lambda$ increases. Eq.~\eqref{eq.critical31contra} indicates order-disorder phase transition for all values of $\lambda > 0$. The critical point $p_c(\lambda)$ also separates between order and disorder phases, as exhibited in Fig.~\ref{fig.31contra}(b).

\subsubsection{Scenario three-one with independence agents}
\begin{figure}[t]
	\centering
	\includegraphics[width=8.5cm]{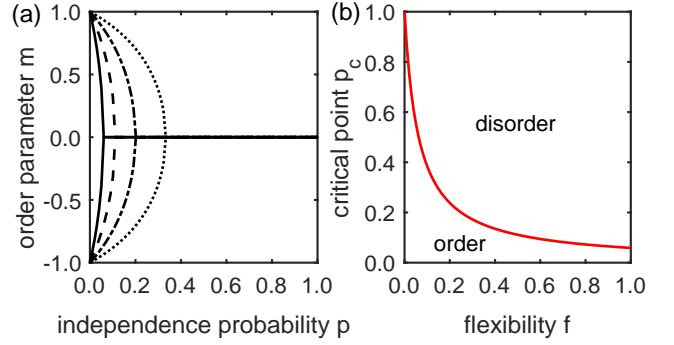}
	\caption{(a) Order parameter $m$ versus independence probability $p$ showing second-order phase transitions of the scenario three-one with independence agents [Eq.~\eqref{eq.orderindep31}] for typical values of flexibility $f$ (solid line: $f = 1$, dashed line: $f= 1/2$, dash-dotted line: $f= 1/4$, dotted line: $f= 1/8$). (b) Critical point $p_c$ versus flexibility $f$ according to Eq.~\eqref{critical-point3_indep}.  The critical point $p_c$ decreases exponentially as the flexibility $f$ increases and it also separates order-disorder phase of the system.}
	\label{fig.31indep}
\end{figure}

For scenario three-one with independence agents, eight agent combinations follow the conformity rule [ illustration~\eqref{eq.contra31}] and eight other agent combinations follow the independence rule [illustration~\eqref{eq.indep31}]. Eight agent combinations make the agents change their opinion/state from 'up' to 'down', and vice versa. Therefore, the probability density of agent-up state increases $\gamma_{+}$ and decreases $\gamma_{-}$ are given by
\begin{equation}\label{gamma3_indep}
	\begin{aligned}
		\gamma_{+}(c)  = & \, 4\,f\,p \left( 1-c \right) +3\,{c}^{2} \left( 1-c \right) ^{2} \left( 1-
		p \right) \\
		&+{c}^{3} \left( 1-c \right)  \left( 1-p \right), \\
		\gamma_{-}(c) = & \, 4\,f\,p\,c+3\,{c}^{2} \left( 1-c \right)^{2} \left( 1-p \right) \\
		&+ c
		\left( 1-c \right) ^{3} \left( 1-p \right).
	\end{aligned}
\end{equation}
Again, for a stationary condition $\gamma_{+} = \gamma_{-}$,  by using the relation $m=2\,c-1$, the order parameter $m$ of the system is given by:
\begin{equation}\label{eq.orderindep31}
	m = \pm \left(\dfrac{1-\left(1+16\,f\right)\,p}{1-p}\right)^{1/2}.
\end{equation}
Based on Eq.~\eqref{eq.orderindep31}, there are several second-order phase transitions for typical values of flexibility $f$, as shown in Fig.~\ref{fig.31indep}(a). The critical point $p_c$ also decreases exponentially as $f$ increases
\begin{equation}\label{critical-point3_indep}
	p_c = \dfrac{1}{1+16\,f},
\end{equation}
which indicates the  majority-minority opinion coexist for $p < p_c(f)$ and the stalemate situation for $p > p_c(f)$ for all values of $f > 0$. The situation is similar to scenario two-two with independence agents, i.e., for small flexibility factor $f$, the probability to reach a stalemate situation is high, and vice versa.  When $\lambda = f$, we find that $p_c$ of scenario three-one with independence agents is smaller than $p_c$ of scenario two-two with contrarian agents. 

\subsection{Probability density function}

In the previous part, we have shown that the model with independence agents undergoes a second-order phase transition, with the critical point depending on the flexibility factor $f$ for the system with independence agents. In this part, we analyze the phase transition of the model from the probability density function $P(m,t)$ of the order parameter $m$ at time $t$. For simplicity without losing generality, here, we only consider the model with independence agents.

\subsubsection{Scenario two-two with independence agents}
The probability density function $P(m,t)$ of the order parameter $m$ at time $t$, for a large system $(1 < N \ll \infty)$,  can be described approximately by the one-dimensional Fokker-Planck equation as follows~\cite{frank2005nonlinear}:
\begin{equation}\label{eq.fokker-planck}
	\dfrac{N}{4}\dfrac{\partial}{\partial t} P(m,t) = \dfrac{\partial^2}{\partial m^2} (\xi_1 P(m,t)) -\dfrac{N}{2} \dfrac{\partial }{\partial m} (\xi_2 P(m,t)),
\end{equation}
where $\xi_1$ and $\xi_2$ are the diffusion and drift coefficients, respectively, defined as:
\begin{equation}\label{diffus-drift_coef}
	\begin{aligned}
		\xi_1 &= \left(\gamma_{+}(m,N)+\gamma_{-}(m,N)\right)/2, \\
		\xi_2 & = \gamma_{+}(m,N)-\gamma_{-}(m,N).
	\end{aligned}
\end{equation}
For a stationary condition, Eq.~\eqref{eq.fokker-planck} has a general solution as follows:
\begin{equation}\label{fokker-planck_solution}
	P(m) = \dfrac{K}{\xi_1}\exp\int \dfrac{N\xi_2}{2\, \xi_1} dm,
\end{equation}
where $K$ is normalization constant that satisfies $\int_{+1}^{-1} P(m,t) \,dm = 1$.

We obtain the diffusion $\xi_1 $ and drift $\xi_2$ coefficients of the scenario two-two with independence agents from Eqs.~\eqref{gamma2_independent} and \eqref{diffus-drift_coef}:
\begin{equation}\label{diffus-drift_2nd}
	\begin{aligned}
		\xi_1& =\left(\left( {m}^{2}+4\,f-1 \right) p-{m}^{2}+1\right)/4, \\
		\xi_2&= \left(\left(m^3-\left(1+4\,f\right)m\right)p-m^3+m\right)/2,
	\end{aligned}
\end{equation}
and from Eqs.~\eqref{fokker-planck_solution} and \eqref{diffus-drift_2nd}, the probability density function $P(m)$ of the system is given by:
\begin{equation}\label{eq.probdens22indep}
	\begin{aligned}
	&P(m)  \sim \left(\left(m^2+4\,f-1\right)p-m^2+1\right)^{-1}  \\ 
		 &\times \exp\left(\frac{N \left(4pf \ln \left(m^2+4f-1\right)p-m^2+1\right)}{\left(1-p\right)}+\frac{m^2N}{2} \right).
	\end{aligned}
\end{equation}

\begin{figure}[t!]
	\centering
	\includegraphics[width=8.5cm]{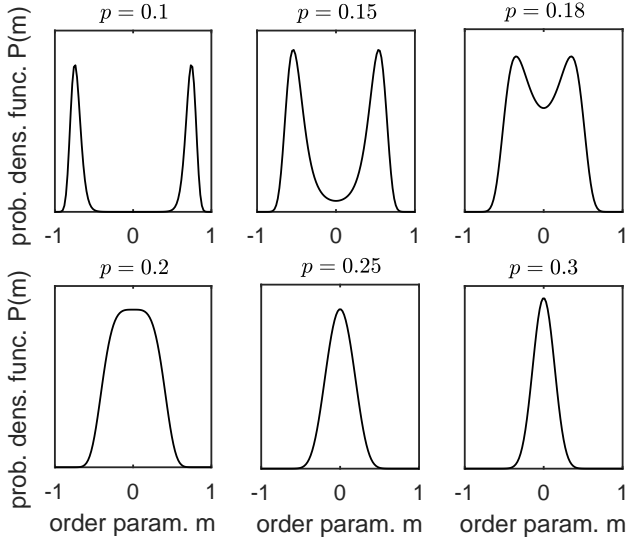}
	\caption{Probability density function $P (m)$ plotted versus order parameter $m$ for scenario two-two involving independence agents. In this plot, we set the population number $N = 200$ and the flexibility factor $f = 1$.  The opinion polarization appears for small values of independence probability $p$, indicating the existence of a majority opinion. By increasing $p$, the system moves towards the non-polarized states, which form a single maximum at $m=0$, indicating no majority opinion in the system. This phenomenon is a typical second-order phase transition.}
	\label{fig.probdens22indep}
\end{figure}

Plot of Eq.~\eqref{eq.probdens22indep} for typical values of $p$, $f = 1$, and $N = 200$ is shown in Fig.~\ref{fig.probdens22indep}. For small probability independence $p$, there is a polarization with $P(m)$ maximum at $\pm m(N)$ and it indicates that there is a majority opinion in the system. When $p$ increases, the polarization approaches each other and makes a single maximum at $m = 0$. In other words, the system goes toward the nonpolarized state, which means that there is no majority opinion. This phenomenon indicates a typical second-order phase transition. From a sociophysics point of view, for $p = 0$, all members have the same opinion indicated by the order parameter $m = \pm 1$. In other words, the system is in complete consensus. The consensus decreases until the independence probability $p =p_c$, and for $p>p_c$ the system is in a stalemate situation.

\subsubsection{Scenario three-one with independence agents}
\begin{figure}[t]
	\centering
	\includegraphics[width=7.8 cm]{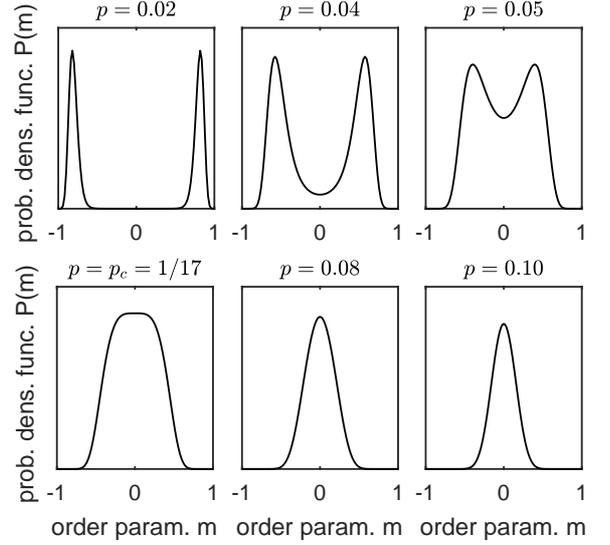}
	\caption{Probability density function $P (m)$ plotted versus order parameter $m$ for scenario three-one involving independence agents with $N = 200$ and $f = 1$ (same parameters as in Fig.~\ref{fig.probdens22indep}).  Similar to Fig.~\ref{fig.probdens22indep}, the result for this scenario also exhibits a second-order phase transition.}
	\label{fig.prob31}
\end{figure}

The diffusion coefficient $\xi_1$ and drift $\xi_2$ from this system can be obtained from Eqs.~\eqref{gamma3_indep} and \eqref{diffus-drift_coef}: 
\begin{equation}\label{diffus-drift3}
	\begin{aligned}
		\xi_1 =& \left(\left(1-p\right)\left(m^4-3\,m^2+2\right)+16\,p\,f\right)/8, \\
		\xi_2 =& \left(\left(m^3-\left(1+16\,f\right)m\right)p-m^3+m\right)/4.
	\end{aligned}
\end{equation}
By inserting Eq.~\eqref{diffus-drift3} into Eq.~\eqref{fokker-planck_solution} and integrating it, the probability density function $P(m)$ of the system can be obtained as follows:
\begin{equation}\label{prob-density}
	\begin{aligned}
		P(m) \sim &  \left(\left(1-p\right)\left(m^4-3\,m^2+2\right)+16\,p\,f\right)^{-1}\\
		&\times \exp \Bigg[ -N\tan^{-1} \left(  \sqrt{\dfrac{p-1}{64pf+p-1}}\left(2m^2-3\right)\right) \\
		& \times \dfrac{\left(32\,p\,f-p+1\right)}{2\sqrt{\left(64\,p\,f+p-1 \right)\left(p-1\right)}} \Bigg] \\ 
		& \times \exp\left(-\frac{N}{4} \ln \left((m^4-3m^2+2)  \left(p-1\right)-16\,p\,f\right)\right).
	\end{aligned}
\end{equation}
Plot of Eq.~\eqref{prob-density} for typical values of  $p$, $f = 1$, and $N = 200$ is shown in Fig.~\ref{fig.prob31}. The result is similar to scenario two-two with the agent of independence that the system undergoes a second-order phase transition. For small values of $p$, the system is in the polarized state, which corresponds to the existence of the majority opinion. As $p$ increases, the system goes toward the unpolarized state with a single maximum at $m=0$.

\subsection{Landau approach for phase transition}

In his theory~\cite{landau1937theory}, Landau assumed that the Gibbs free energy depends not only on the thermodynamics parameter such as temperature and pressure but also on the order parameter such as in Eq.~\eqref{order-param}. Generally, one defines the Landau potential as a function of order parameter $m$ and any parameter describing the system state, e.g., in this case, the probability of independence $p$. Therefore, to analyze the phase transition in this system, the first three terms of the simplified Landau potential can be written as follows:
\begin{equation}\label{landau-potential}
V(m,p) = V_0 + V_1 \, m^2 + V_2 \, m^4,
\end{equation}
where the parameters $V_1$ and $V_2$, in general, can be as a function of contrarian or independence probability $p$. Based on Eq.~\eqref{landau-potential}, the phase transition occurs when $V_1=0$ and $V_2>0$. 

\begin{figure}[t]
	\centering
	\includegraphics[width=8.5cm]{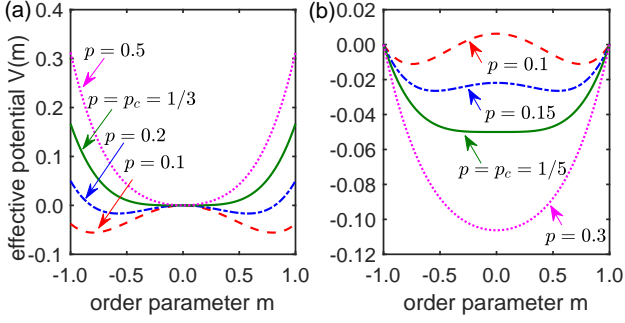}
	\caption{The effective potential $V(m)$ versus order parameter $m$ based on the Sznajd model on a complete graph for (a)~scenario two-two with contrarian agents and (b) scenario two-two with independence agents.  Below the critical point $p_c = 1/3$ ($\lambda = 1$), the system with contrarian agents shown in panel (a) is in bi-stable states, which correspond to two meta-stable states. The bi-stable states also appear in case (b) with independence agents below the critical point $p_c = 1/5$ ($f = 1$). For both scenarios, the stability decreases as $p$ increases and moves towards the mono-stable state at $m = 0$. This situation also indicates that both scenarios undergo a second-order phase transition.}
	\label{bis-potential}
\end{figure}

To obtain the 'effective potential' in this system, firstly, we define the 'effective force' that is, the difference between the probability density of spin up increasing $\gamma_{+}$ and decreasing $\gamma_{-}$, $f = \gamma_{+}- \gamma_{-}$ \cite{nyczka2012opinion}. Therefore, the effective potential of the system is obtained by using $V(m) = -\int f(m) \, dm$. For the scenario two-two with contrarian agents, its effective potential is given by
\begin{align}\label{eff-potential}
V(m) = \left(p\,(1+2\lambda)-1 \right)m^2/4+\left(1-p\,(1-2\lambda)\right)m^4/8
\end{align}
Therefore, from Eqs.~\eqref{landau-potential} and \eqref{eff-potential}, the parameters $V_1$ and $V_2$ are
\begin{align}\label{parameter-AB}
		V_1= & \left(p\,(1+2\lambda)-1 \right)/4,\\
		V_2 = &\left(1-p\,(1-2\lambda)\right)/8,
\end{align}
where the critical point corresponding to $V_1 = 0$, i.e.,
\begin{equation}\label{eq.30}
	p_c = \dfrac{1}{1+2\lambda},
\end{equation}
is the same as Eq.~\eqref{critical-point22contra}. We can write $V_2(\lambda) = \lambda/\left(2+4\lambda\right)>0$ for all values of $\lambda$. Plot of Eq.~\eqref{eff-potential} for several values of contrarian probability $p$ is shown in Fig.~\ref{bis-potential}(a).

For scenario two-two with independence agents, the effective potential is given by
\begin{equation}\label{eff-potential_2indep}
    \begin{aligned}
	V(m) =& \, \Big( \left(1-\left(1+8\,f\right)p\right)-\left(2-\left(2+8\,f\right)p\right)m^2 \\
	&+\left(1-p\right)m^4 \Big)/16,
    \end{aligned}
\end{equation}
where the parameters $V_1$ and $V_2$ are given by:
\begin{align}
		V_1 =&  -\left(1-\left(1+4\,f\right)p\right)/8\\
		V_2 =& \left(1-p\right)/16  = f/ (4+16\,f) >0.
\end{align}
Note that $V_2$ is always positive for all values of $f$.  The critical point corresponding to $V_1 = 0$ is $p_c = 1/(1+4\,f)$, which is consistent with Eq.~\eqref{critical-point_indep22}.  Plot of Eq.~\eqref{eff-potential_2indep} for several values of probability independence $p$ is given in Fig.~\ref{bis-potential}(b).

For scenario three-one with contrarian agents, its effective potential (not plotted) is given by
\begin{equation}\label{eff-potential3_contra}
    \begin{aligned}
	V(m) =& \, \Big(\left(1-\left(1+\lambda\right)p\right)5\,m^4-\left(1-\left(1+5\lambda\right)p\right)10\,m^2 \\
	&+\left(5-\left(5+13\lambda\right)p\right)\Big)/160,
	\end{aligned}
\end{equation}
where the parameters $V_1$ and $V_2$ are given by
\begin{align}\label{eq.param-V_contra}
		V_1 & = -\left(1-\left(1+5\lambda\right)p\right)/16, \\
		V_2 & = \left(1-\left(1+\lambda\right)p \right)/32 = \lambda/(8+40\,\lambda) > 0.
\end{align}
The critical point (when $V_1 = 0$) is $p_c =1/(1+5\,\lambda)$, which confirms Eq.~\eqref{eq.critical31contra}. The effective potential for the scenario three-one with independence agents (not plotted) is given by:
\begin{align}\label{eff-potential3_indep}
	V(m) =~&\Big(\left(1-\left(1-32\,f\right)p\right)
	       -\left(2-\left(2+32\,f\right)p\right)m^2\nonumber\\
	      &+\left(1-p\right)m^4\Big)/32,
\end{align}
where the parameters $V_1$ and $V_2$ are given by
\begin{align}\label{param-AB3_indep}
		V_1 =& -\left(1-\left(1+16f\right)p\right)/16, \\
		V_2 =& (1-p)/32 = f/(2+32\,f)>0.
\end{align}
The critical point in this case is $p_c = 1/(1+16\,f)$. 

For all scenarios, based on the parameters $V_1$ and $V_2$,  the system undergoes a second-order phase transition for all values of $\lambda$ and $f$. As exhibited in Figs.~\ref{bis-potential}(a) and~\ref{bis-potential}(b) for scenario two-two and for $p$ below the critical point, the potential is in a bi-stable state indicating an ordered phase. As $p$ increases, the potential starts reaching a minimum and towards monostable at $m=0$, indicating a new disordered phase. This phenomenon is a typical second-order phase transition. We will obtain the same phenomenon also for scenario three-one.

\subsection{Numerical simulation}
We perform numerical simulations for $p \in [0,1]$ for all scenarios to estimate the critical point of the model. The relevant control parameter is a ratio between contrarian or independence probability with conformity probability $r = p_2/p_1$. We vary the total population $N$, e.g., $N = 256, 512, 1024, 2048, 8192$ and using the finite-size relation [Eqs. \eqref{eq4}--\eqref{eq7}] to estimate the critical point and the critical exponents of the system. The initial condition is $0.5$ (disorder state).

\begin{figure*}[t]
	\centering
	\includegraphics[width=15cm]{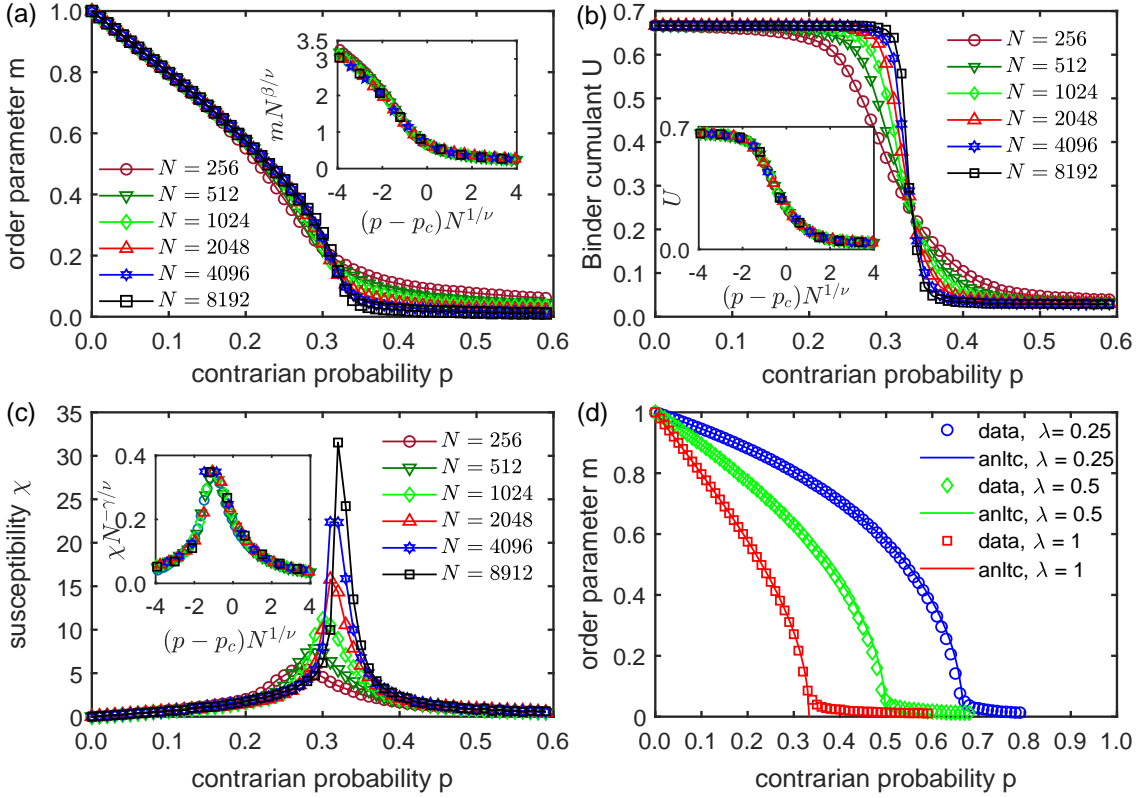}
	\caption{(a)-(c) Numerical results of order parameter, Binder cumulant, and susceptibility for the scenario two-two with contrarian agents based on the outflow dynamics with $\lambda = 1$ on a complete graph. The critical point can be obtained from the lines of intersection of Binder cumulant $U$ versus probability of contrarian behavior $p$ in (b). The best data collapses are obtained for $p_c \approx 0.33$, $\beta \approx 0.5$, $\gamma \approx 1$, and $\nu \approx 2$, indicating this model is in the same universality class as the mean-field Ising model. Panel (d) shows that the numerical results with $N = 10000$ and several values of $\lambda$ are in good agreement with the analytical results [Eq.~\eqref{critical-point22contra}].}
	\label{fig.num22contra}
\end{figure*}

\subsubsection{Scenario two-two with contrarian agents}

The parameter control for this scenario is $r = \lambda\,p / (1-p)$; therefore, the real control parameter only depends on the contrarian factor $\lambda$ and contrarian probability $p$. The order parameter $m$ versus the contrarian probability $p$ is shown in Fig.~\ref{fig.num22contra}(a), where the inset graph shows the best collapse for all values of $N$. Based on Ref.~\cite{binder1981finite}, the critical point can be obtained from the intersection of lines between Binder cumulant $U$ versus $p$, that occurred at $p = p_c \approx 0.33$ for $\lambda = 1$ as shown in Fig.~\ref{fig.num22contra}(b). As shown in Fig.~\ref{fig.num22contra}(c), the ''peak" of the susceptibility $\chi$ shifts towards the critical point $p_c \approx 0.33$ for $N$ increases. This result confirms the analytical result in Eq.~\eqref{order-parameter2nd}, in which for  $m = 0$ and $\lambda = 1$, we find $p = p_c = 1/3$. We also estimate the critical exponents $\beta, \gamma, $ and $\nu$ using the finite-size scaling relations in Eqs.~\eqref{eq4}--\eqref{eq7} and find that the critical exponents that make all the values of $N$ collapse are $\beta \approx 0.5, \gamma \approx 1,$ and $\nu \approx 2$. These exponents are universal, i.e., we obtain the same values of $\beta, \nu,$ and $\gamma$ for all values of $N$. Based on the values of the critical exponents, our results indicate that this model is in the same universality class as that of the kinetic exchange opinion model with two-one agent interactions~\cite{crokidakis2014phase, crokidakis2012role, biswas2009model}, as well as that of the mean-field Ising model. The numerical estimate for the critical exponent $\beta = 1/2$ also agrees with Eq.~\eqref{order-parameter2nd} which can be written in form of $m \sim 
(p-p_c)^{\beta}$ for all values of $\lambda$. 

Equation~\eqref{critical-point22contra} is also confirmed numerically, as shown in Fig.~\ref{fig.num22contra}(d), indicating that there are phase transitions for $\lambda > 0$.  One can see the good agreement between the analytical and simulation results. The critical probability of contrarian $p$ decreases as $\lambda$ increases. For high $\lambda$, a spin-flip occurs more frequently in the contrarian case and makes the system more disordered. From a sociophysics point of view, at $ p = 0 $, a complete consensus is reached with all members having the same opinion. $p <p_c$ indicates a majority opinion exists, while $p \geq p_c$ indicates no majority opinion or the system is in a stalemate situation. Furthermore, in a society with high contrarian behavior, the possibility to achieve a status quo or a stalemate situation is relatively high.

\subsubsection{Scenario two-two with independence agents}

\begin{figure}[t]
	\centering
	\includegraphics[width = 7 cm]{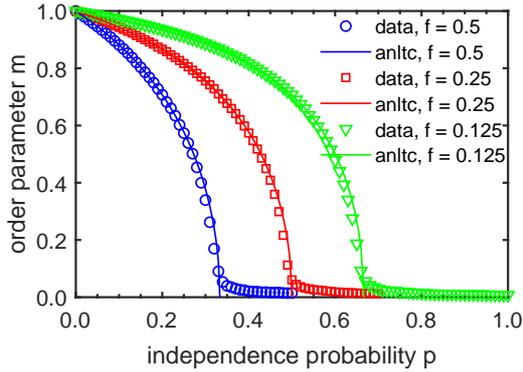}
	\caption{Order parameter $m$ versus probability of independence $p$ for typical values of $f$. Symbols represent the numerical results for the population $N = 10000$, while solid lines represent the analytical result from Eq.~\eqref{order-param2_indep}.}
	\label{fig.orderindep22}
\end{figure}

The control parameter for this scenario is $r = fp/(1-p)$ and we only prove the order parameter $m$ in Eq.~\eqref{order-param2_indep}, i.e., there are several phase transitions for all values of inflexibility ($f > 0$). We use $N = 10000$ and average over $1000$ simulation for each point as exhibited in Fig.~\ref{fig.orderindep22}. It can be seen that the numerical results for several values of $f$ also agree with the analytical result. From a sociophysics point of view,  majority-minority opinion stills exist below the critical point $p_c(\lambda)$, while for $p > p_c(\lambda)$ there is no majority opinion or the systems in a stalemate situation. 

One can see that the critical probability of independence decreases as $f$ increases. For high flexibility $f$, a spin-flip occurs more frequently in the independence case and makes the system more chaotic. Therefore, the probability of the system reaching the stalemate situation is higher. In other words, for high $f$, the possibility of consensus is tiny.

\subsubsection{Scenario three-one with contrarian agents}

\begin{figure*}[t]
	\centering
	\includegraphics[width = 14cm]{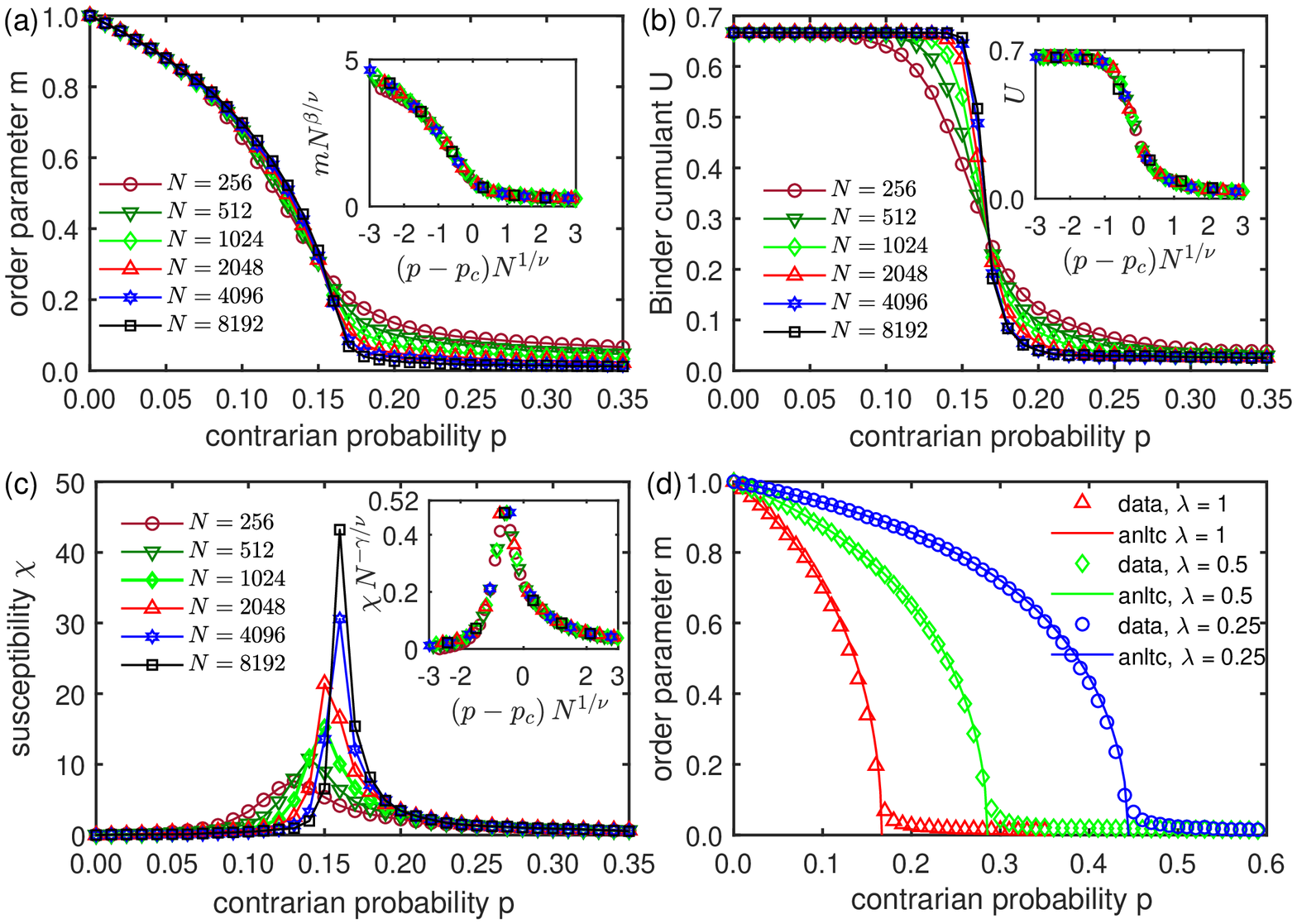}
	\caption{(a)--(c) Numerical results of order parameter, Binder cumulant, and susceptibility for the scenario three-one with contrarian agents based on the outflow dynamics with $\lambda = 1$ on a complete graph. The critical point can be obtained from the lines of intersection of Binder cumulant $U$ versus probability of contrarian behavior $p$ in (b). The best data collapses are obtained for $p=p_c \approx 0.166$, $\beta \approx 0.5$, $\gamma \approx 1$, and $\nu \approx 2$, indicating this model is in the same universality class as the mean-field Ising model. Panel (d) shows that the numerical results with $N = 10000$ and several values of $\lambda$ are in good agreement with the analytical results [Eq.~\eqref{order-parameter3_contra}].}
	\label{fig.numcontra31}
\end{figure*}

The numerical results for the scenario three-one with contrarian agents are given in Figure \ref{fig.numcontra31}. The order parameter $m$ versus the contrarian probability $p$ is shown in Fig.~\ref{fig.numcontra31}(a). We also find the critical point $p_c$ from the intersection of lines between Binder cumulant $U$ versus probability $p$ that occurred at $p=p_c \approx 0.166$ for $\lambda = 1$ as shown in Fig.~\ref{fig.numcontra31}(b). The "peak" of the susceptibility $\chi$ shifts towards the critical point $p_c \approx 0.166$ for $N$ increases as shown in Fig.~\ref{fig.numcontra31}(c). We also find the critical exponents $\beta \approx 0.5, \gamma \approx 1,$ and $\nu \approx 2$ using finite-size scaling \eqref{eq4}-\eqref{eq7} (inset graphs). From this results, the model is also found to be in the same universality class as the mean-field Ising model. The numerical estimate for $\beta \approx 0.5$ agrees with Eq.~\eqref{order-parameter3_contra}, which can be written in form $m \sim (p-p_c)^{\beta} = (p-p_c)^{1/2}$ for all values of $\lambda$.  The numerical result of order parameter $m$ for typical values of $\lambda$ also agrees with the analytical result \eqref{order-parameter3_contra} as exhibited in Fig.~\ref{fig.numcontra31}(d). One can see that there are several second-order phase transitions for typical values of $\lambda > 0$. 

\begin{figure}[t]
	\centering
	\includegraphics[width = 7 cm]{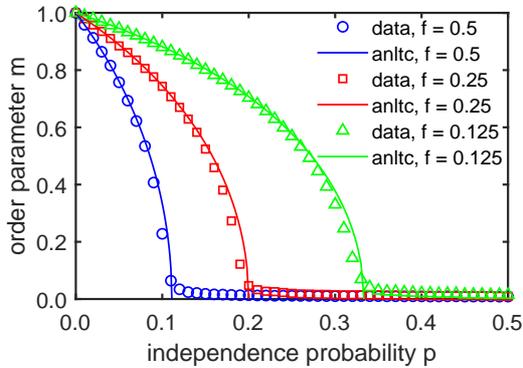}
	\caption{Order parameter $m$ versus probability of independence $p$ for typical values of $f$. Symbols represent the numerical results for population $N = 10000$, while solid lines represent the analytical results from Eq.~\eqref{eq.orderindep31}.}
	\label{fig.numorder31indep}
\end{figure}

\subsubsection{Scenario three-one with independence agents}

In this scenario, we also performed numerical simulation only for order parameter $m$ with typical flexibility value $f$. We use the population $N = 10000$ and average over $1000$ simulations for each point as exhibited in Fig.~\eqref{fig.numorder31indep}. Numerical and analytical results are in agreement \eqref{eq.orderindep31} for typical values of $f$. We also compare scenario three-one with scenario two-two. The critical point $p_c(f)$ is smaller in scenario three-one than in scenario two-two for the same value of $f$. The same results are obtained in both the scenario with independence agents and the scenario with contrarian agents due to the higher number of persuaders will cause the system to be more likely in a stalemate situation. 

From the analytical and simulation results, for the same scenario, the critical point for the same value of $\lambda = f$ is more significant for the scenario with contrarian agents than the scenario with independence agents because independence behavior in the system makes it more chaotic than contrarian behavior in the system. This situation corresponds to the typical independence behavior that involves actions independent from the group norm. In contrast, the spin with the contrarian behavior flips in a more organized manner because it follows the group norm.

\subsubsection{Remarks on the novelty of this study}
Based on the analytical and simulation results, all four  scenarios discussed above actually follow the same dynamical equation:
\begin{equation}
    \dfrac{dm}{dt} = k_1 m - k_2 m^3.
\end{equation}
This indicates all four scenarios undergo second-order phase transition, in which $m$ is the order parameter and the $k_1$, $k_2$ parameters depend on the contrarian or independence probability $p$, the flexibility $f$ and the level of contrarian $\lambda$. For example, the order parameter $m$ is explicitly described by Eqs.~\eqref{order-parameter2nd} and \eqref{order-param2_indep} for scenarios two-two with contrarian agents and independence agents, respectively. Meanwhile, for scenarios three-one with contrarian agents and independence agents, the order parameter $m$ is explicitly described by Eqs.~\eqref{order-parameter3_contra} and \eqref{eq.orderindep31}, respectively.

The critical points $p_c$ of the scenario with contrarian and independence agents depend on the level of ``noise parameter" $\lambda$ and $f$, respectively [c.f. Eqs.~\eqref{critical-point22contra} and \eqref{critical-point_indep22} for scenario two-two; and Eqs.~\eqref{eq.critical31contra} and \eqref{critical-point3_indep} for scenario three-one]. Interestingly, although they have different microscopic interactions, even have a different type of ``noises" $\lambda$ or $f$, but based on our numerical simulations, all four scenarios have the same critical exponents $\beta \approx 0.5, \nu \approx 2.0$ and $\gamma \approx 1.0$, indicating that that they are identical and in the same universality class as the mean-field Ising model.  This argument can be considered as the novelty of this work.

\section{Summary and outlook}
\label{sec.4}
We have investigated the outflow dynamics or the Sznajd model for four-agent (four-spin) local interaction with two different scenarios on a complete graph. In the first scenario, two agents persuade the other agents wherever in population; meanwhile, in the second scenario, three agents persuade the fourth agent wherever in population. For each scenario, we considered two types of social behaviors, namely contrarian and independence. These social behaviors act like noises that cause the system to undergo an order-disorder phase transition. We analyze the effect of both social behaviors on the phase transition of the systems and compare the results.

Based on the calculations of phase transition and universality of the model, we found that all systems undergo second-order phase transition for all values of contrarian factor $\lambda$ and flexibility $f$. The critical point $p_c$ depends on the contrarian factor $\lambda$ or on the flexibility $f$, where $p_c$ decreases exponentially as $\lambda$ or $f$ increases. For high-level contrarian and independence factors (nonconservative society), the possibility to reach a consensus is small. Otherwise, the possibility of reaching a status quo or a stalemate situation is high. We also found that the critical point for scenario three-one is smaller than scenario two-two for both contrarian and independence cases for the same value of $\lambda $ or $f$. In addition, the critical probability $p_c$ is smaller in the scenario with independence agents than the contrarian for $\lambda = f$. Using finite-size scaling relations, the critical exponents for both scenarios with contrarian agents are $\beta \approx 0.5, \gamma  \approx 1$, and $\nu \approx 2$.  Therefore, our results suggest that the model is in the same universality class as the mean-field Ising model.

From this study, we suggest that the existence of a group with independent behaviors may disrupt the social cohesion of a society more than a group of people with a tendency of contrarian behaviors. From either the two-two or three-one scenario of our result, the existence of a significant group with independent opinions in society makes reaching a consensus harder than that with contrarian behavior. The difficulty of reaching a consensus amid socio-political disturbance is a formula for creating an unstable society. Thus, this model corroborates the fact that the existence of a system with strong duo political-social entities that are contrary to each other in many issues can be more stable than a single omnipotent political-social entity with many unorganized independent movements, especially during strong socio-political turbulence.  

\section*{Data Availability}
The raw/processed data required to reproduce these findings are available to download from \verb|https://github.com/muslimroni/Sznajd2231|.

\section*{CRediT authorship contribution statement}
\textbf{R. Muslim:} Conceptualization, Methodology, Software, Formal analysis, Investigation, Data Curation, Writing - original draft, Visualization. \textbf{M.J. Kholili:} Validation, Formal analysis, Writing - review \& editing. \textbf{A.R.T. Nugraha:}  Writing - review \& editing, Supervision, Funding acquisition. 

\section*{Declaration of Interests}
The authors declare that they have no known competing financial interests or personal relationships that could have appeared to influence the work reported in this paper.

\section*{Acknowledgments}
R.M. is supported by postdoctoral fellowship under LIPI/BRIN talent management program. We acknowledge Dr. Rinto Anugraha NQZ from Gadjah Mada University for his guidance during the graduate study of R.M.


\begin{thebibliography}{10}
\expandafter\ifx\csname url\endcsname\relax
  \def\url#1{\texttt{#1}}\fi
\expandafter\ifx\csname urlprefix\endcsname\relax\def\urlprefix{URL }\fi
\expandafter\ifx\csname href\endcsname\relax
  \def\href#1#2{#2} \def\path#1{#1}\fi

\bibitem{castellano2009statistical}
C.~Castellano, S.~Fortunato, V.~Loreto, Statistical physics of social dynamics,
  Rev. Mod. Phys. 81 (2009) 591.

\bibitem{serge2016sociophysics}
S.~Galam, Sociophysics: A Physicist's Modeling of Psycho-political Phenomena,
  Springer, Boston, MA, 2016.

\bibitem{sen2014sociophysics}
P.~Sen, B.~K. Chakrabarti, Sociophysics: an introduction, Oxford, Oxford
  University Press, 2014.
  
\bibitem{javarone2014network}
M.~A.Javarone, Network strategies in election campaigns . Journal of Statistical Mechanics: Theory and Experiment 8, (2014) 08013.

\bibitem{stauffer2009encyclopedia}
D.~Stauffer, Phase transitions on fractals and networks, in: Encyclopedia of
  Complexity and Systems Science, Berlin, Springer, 2007, pp. 193--221.

\bibitem{myer2013}
D.~G. Myers, Social Psychology, New York, McGraw Hill, 2013.

\bibitem{sznajd2000opinion}
K.~Sznajd-Weron, J.~Sznajd, Opinion evolution in closed community, Int. J. Mod.
  Phys. C 11 (2000) 1157--1165.

\bibitem{galam2008sociophysics}
S.~Galam, Sociophysics: A review of \text{Galam} models, Int. J. Mod. Phys. C
  19 (2008) 409--440.

\bibitem{liggett1985interacting}
T.~M. Liggett, Interacting particle systems, Berlin, Springer, 1985.

\bibitem{mobilia2003majority}
M.~Mobilia, S.~Redner, Majority versus minority dynamics: Phase transition in
  an interacting two-state spin system, Phys. Rev. E 68 (2003) 046106.

\bibitem{galam2002minority}
S.~Galam, Minority opinion spreading in random geometry, Eur. Phys. J. B 25
  (2002) 403--406.

\bibitem{krapivsky2003dynamics}
P.~L. Krapivsky, S.~Redner, Dynamics of majority rule in two-state interacting
  spin systems, Phys. Rev. Lett. 90 (2003) 238701.

\bibitem{biswas2009model}
S.~Biswas, P.~Sen, Model of binary opinion dynamics: Coarsening and effect of
  disorder, Phys. Rev. E. 80 (2009) 027101.

\bibitem{nail2000proposal}
P.~R. Nail, G.~MacDonald, D.~A. Levy, Proposal of a four-dimensional model of
  social response., Psychol. Bull. 126 (2000) 454.

\bibitem{galam2004contrarian}
S.~Galam, Contrarian deterministic effects on opinion dynamics: ``the hung
  elections scenario", Physica A 333 (2004) 453--460.

\bibitem{galam2007role}
S.~Galam, F.~Jacobs, The role of inflexible minorities in the breaking of
  democratic opinion dynamics, Physica A 381 (2007) 366--376.

\bibitem{nyczka2013anticonformity}
P.~Nyczka, K.~Sznajd-Weron, Anticonformity or independence?—insights from
  statistical physics, J. Stat. Phys. 151 (2013) 174--202.

\bibitem{mobilia2003does}
M.~Mobilia, Does a single zealot affect an infinite group of voters?, Phys.
  Rev. Lett. 91 (2003) 028701.

\bibitem{willis1963two}
R.~H. Willis, Two dimensions of conformity-nonconformity, Sociometry (1963)
  499--513.

\bibitem{willis1965conformity}
R.~H. Willis, Conformity, independence, and anticonformity, Hum. Relat. 18
  (1965) 373--388.

\bibitem{macdonald2004expanding}
G.~MacDonald, P.~R. Nail, D.~A. Levy, Expanding the scope of the social
  response context model, Basic Appl. Soc. Psych. 26 (2004) 77--92.

\bibitem{nail2011development}
P.~R. Nail, G.~MacDonald, On the development of the social response context
  model, in: The science of social influence: Advances and future progress, New
  York, Psychology Press, 2007, pp. 193--221.
  
\bibitem{javarone2014social}
M.~A.Javarone, Social influences in opinion dynamics: the role of conformity, Physica A: Statistical Mechanics and its Applications 414 (2014) 19-30.

\bibitem{javarone2015conformism}
M.~A.Javarone, T.~ Squartini, Conformism-driven phases of opinion formation on heterogeneous networks: the q-voter model case, Journal of Statistical Mechanics: Theory and Experiment, 10 (2015) P10002.

\bibitem{de2005spontaneous}
M.~S. de~la Lama, J.~M. L{\'o}pez, H.~S. Wio, Spontaneous emergence of
  contrarian-like behaviour in an opinion spreading model, Europhys. Lett. 72
  (2005) 851.

\bibitem{calvelli2019phase}
M.~Calvelli, N.~Crokidakis, T.~J.~P. Penna, Phase transitions and universality
  in the sznajd model with anticonformity, Physica A 513 (2019) 518--523.
  
\bibitem{muslim2020phase}
R.~Muslim, R.~Anugraha, S.~Sholihun, M.~F.~Rosyid, Phase transition of the Sznajd model with anticonformity for two different agent configurations, Int. J. Mod. Phys. C 31 (2020) 2050052.

\bibitem{sznajd2011phase}
K.~Sznajd-Weron, M.~Tabiszewski, A.~M. Timpanaro, Phase transition in the
  sznajd model with independence, Europhys. Lett. 96 (2011) 48002.

\bibitem{crokidakis2015inflexibility}
N.~Crokidakis, P.~M.~C. de~Oliveira, Inflexibility and independence: Phase
  transitions in the majority-rule model, Phys. Rev. E 92 (2015) 062122.

\bibitem{chmiel2015phase}
A.~Chmiel, K.~Sznajd-Weron, Phase transitions in the q-voter model with noise on a duplex clique, Phys. Rev. E, 92 (2015) 052812.

\bibitem{abramiuk2019independence}
A.~Abramiuk, J.~Paw{\l}owski, K.~Sznajd-Weron, Is independence necessary for a discontinuous phase transition within the q-voter model?, Entropy 21 (2019) 521.

\bibitem{nowak2021discontinuous}
B.~Nowak, B.~Stoń, K.~Sznajd-Weron, Discontinuous phase transitions in the multi-state noisy q-voter model: quenched vs. annealed disorder, Sci. Rep., 11 (2021) 1--13.

\bibitem{civitarese2021external}
J.~Civitarese, External fields, independence, and disorder in q-voter models, Phys. Rev. E, 103 (2021) 012303.

\bibitem{hofstede2001}
G.~Hofstede, Culture's consequences: Comparing values, behaviors, institutions and organizations across nations, California, Sage publications, 2001.

\bibitem{hofstede2010}
G.~Hofstede, G.~J. Hofstede, M.~Minkov, Cultures and organizations: Software of
  the mind, New York, McGraw Hill, 2010.

\bibitem{solomon2010consumer}
M.~R. Solomon, G.~Bamossy, S.~Askegaard, M.~K. Hogg, Consumer behavior: a
  European perspective 4th edn, England, Pearson Education, 2010.

\bibitem{hofstedeIDV}
See \url{https://www.hofstede-insights.com/country-comparison} for further
  details about IDV of various countries in the world.

\bibitem{nyczka2012phase}
P.~Nyczka, K.~Sznajd-Weron, J.~Cis{\l}o, Phase transitions in the q-voter model
  with two types of stochastic driving, Phys. Rev. E 86 (2012) 011105.

\bibitem{binder1981finite}
K.~Binder, Finite size scaling analysis of ising model block distribution
  functions, Z. Phys. B: Condens. Matter 43 (1981) 119--140.

\bibitem{cardy1996scaling}
J.~Cardy, Scaling and renormalization in statistical physics, Vol.~5,
  Cambridge, Cambridge University Press, 1996.

\bibitem{crokidakis2014phase}
N.~Crokidakis, Phase transition in kinetic exchange opinion models with
  independence, Phys. Lett. A 378 (2014) 1683--1686.

\bibitem{krapivsky2010kinetic}
P.~L. Krapivsky, S.~Redner, E.~Ben-Naim, A kinetic view of statistical physics,
  Cambridge, Cambridge University Press, 2010.

\bibitem{frank2005nonlinear}
T.~D. Frank, Nonlinear Fokker-Planck equations: fundamentals and applications,
  Berlin, Springer, 2005.

\bibitem{landau1937theory}
L.~D. Landau, On the theory of phase transitions, Zh. Eksp. Teor. Fiz. 7 (1937)
  19--32.

\bibitem{nyczka2012opinion}
P.~Nyczka, J.~Cis{\l}o, K.~Sznajd-Weron, Opinion dynamics as a movement in a
  bistable potential, Physica A 391 (2012) 317--327.

\bibitem{crokidakis2012role}
N.~Crokidakis, C.~Anteneodo, Role of conviction in nonequilibrium models of
  opinion formation, Phys. Rev. E 86 (2012) 061127.

\end{thebibliography}

\end{document}